\documentclass[12pt]{article}
\usepackage{eurosym}
\usepackage{amssymb}
\usepackage{amsmath}
\usepackage{cite}

\newtheorem{theorem}{Theorem}

\newtheorem{corollary}[theorem]{Corollary}

\newtheorem{definition}[theorem]{Definition}

\newtheorem{lemma}[theorem]{Lemma}

\input{tcilatex}

\begin{document}

\title{Linearization of Newton's second law}
\author{Andronikos Paliathanasis\thanks{%
Email: anpaliat@phys.uoa.gr} \\
{\ } {\ \textit{Institute of Systems Science, Durban University of
Technology }}\\
{\ \textit{PO Box 1334, Durban 4000, Republic of South Africa}} \\
{\ }\textit{Departamento de Matem\'{a}ticas, Universidad Cat\'{o}lica del
Norte,} \\
\textit{Avda. Angamos 0610, Casilla 1280 Antofagasta, Chile}}
\maketitle

\begin{abstract}
The geometric linearization of nonlinear differential equation is a robust
method for the construction of analytic solutions. The method is related to
the existence of Lie symmetries which can be used to determine point
transformations such that to write the given differential equation in a
linear form. In this study we employ another geometric approach and we
utilize the Eisenhart lift to geometric linearize the Newtonian system
describing the motion of a particle in a line under the application of an
autonomous force. Our findings reveal that for the oscillator, the Ermakov
potential with or without the oscillator term, and the Morse potential,
Newton's second law can be globally expressed in the form of that of a free
particle.\ This study open new directions for the geometric linearization of
differential equations via equivalent dynamical systems.

\end{abstract}

\section{Introduction}

\label{sec1}

Sophus Lie's theory \cite{lie1,lie2,lie3} stands as a powerful mathematical
tool for the analytic treatment of nonlinear differential equations. A Lie
symmetry vector serves as the generator of transformations that leave a
given differential equation invariant. Consequently, there exist invariant
functions, known as Lie invariants, which are invaluable for simplifying
sets of differential equations through reduction \cite%
{ibra,Bluman,Stephani,olver}. The Lie symmetry analysis offers a systematic
approach to studying differential equations and has been instrumental in
deriving numerous significant results. As a result, it has found widespread
use in the analysis of various dynamical systems \cite%
{r1,r3,r5,ao01,ao02,ao03,ao04,ao05,sw3,swr06,ans1,sw1,mel3,ja1,ja2,ja3,bk1,bk2}%
.

An intriguing property of Lie symmetry analysis is its ability to globally
linearize maximally symmetric differential equations \cite{r3,r10}.
Second-order ordinary differential equations that are invariant under the
elements of the $sl(3,R)$ Lie algebra are considered maximally symmetric 
\cite{ao01}. Consequently, there always exists a point transformation, or
change of variables, where the equation can be globally expressed in the
form of a free particle. This linearization process allows for the
determination of conservations and the analytic solution for the original
equation through a simple application of the point transformation. Due to
this significant property, the process of linearization has been extensively
explored in the literature, leading to the establishment of various criteria 
\cite{q1,q2,q3,q4,q5}. Due to the specific property that a family of
polynomial second-order differential equations is related to geodesic
equations in \cite{am1,am2} some geometric criteria have been established
for the linearization, see also the discussion in \cite{am3,am4}. 

In this study we are interested in the linearization properties of the
Newtonian system with equation of motion%
\begin{equation}
\ddot{x}-F\left( x\right) =0.  \label{ns.01}
\end{equation}%
The latter equation describes the motion for a classical particle under the
application of an autonomous force.

The Lie symmetry analysis for equation (\ref{ns.01}) has been performed by
Sophus Lie, and when $F\left( x\right) =F_{1}x+F_{0}$, the differential
equation is maximally symmetric, that is, it admits eight Lie symmetries the
elements of the $sl\left( 3,R\right) $; straightforward equation (\ref{ns.01}%
) is a linear equation \cite{ao03}. For other functional forms of the force $%
F\left( x\right) $ equation (\ref{ns.01}) admits less Lie symmetry vectors.
Indeed, for $F\left( x\right) =\left( \alpha +\beta x\right) ^{n},~n\neq
0,1,-3$ and $F\left( x\right) =e^{\gamma x},~\gamma \neq 0,~$the admitted
Lie symmetries of (\ref{ns.01}) are two, and form the algebra $A_{2}$ in the
Mubarakzyanov Classification Scheme. Moreover, for $F\left( x\right) =\alpha
\left( x+c\right) +\frac{\beta }{\left( x+c\right) ^{3}}$, $\beta \neq 0$
equation (\ref{ns.01}) admits three Lie symmetries which form the
three-dimensional algebra, $A_{3,8}$, which is more commonly known as $%
sl\left( 2,R\right) $. Finally, for arbitrary function $F\left( x\right) $,
equation (\ref{ns.01}) admits a unique symmetry, which defines the
invariance in time translations.

In the following, we introduce a novel approach where we demonstrate that
equation (\ref{ns.01}) can be linearized for specific functions of the
force, thereby breaking the condition of maximally symmetric equations. To
achieve this, we utilize the geometric framework established by L. Eisenhart 
\cite{el1}. We perform a lift in the dimension of equation (\ref{ns.01}),
transforming it into a system expressed as geodesic equations on a new
(pseudo-) Riemannian manifold. This transformation allows the force term $%
F(x)$ to be attributed to the geometric properties of this manifold.

The Eisenhart lift has found application in a range of studies for the
analysis of classical and quantum systems \cite{en1,en2,en3,en4,en5,en5a}.
In a related work \cite{en6}, authors discuss the equivalence of the
oscillator with the free particle using the Eisenhart lift. Building upon
this foundation, we extend these results by presenting a systematic
approach. Specifically, we demonstrate that for specific nonlinear functions 
$F(x)$, the analytic solution of equation (\ref{ns.01}) corresponds to the
solution of the free particle. The outline of the paper is as follows.

In Section \ref{sec2} we present the Eisenhart lift and we define new
higher-order lifts. Section \ref{sec31} includes the main results of this
analysis where we employ results from differential geometry in order to
determine the functional forms of the force term $F\left( x\right) $ where
equation (\ref{ns.01}) can be linearized via the Eisenhart lift. Finally, in
Section \ref{sec4} we draw our conclusions.

\section{Eisenhart lift}

\label{sec2}

In this section, we delve into the Eisenhart lift, a technique that provides
a geometric representation of dynamical systems, allowing them to be
formulated as sets of geodesic equations \cite{el1}. Through the Eisenhart
lift (also known as the Eisenhart-Duval lift), we introduce an equivalent
dynamical system of higher-dimension, that is, we introduce new dependent
variables. The new dimensions allows us to express the original dynamical
system in the equivalent form of geodesic equations for an extended
manifold. The extended manifold possesses a sufficient number of isometries,
associated with conservation laws, which are essential for recovering the
original dynamical system. This procedure is known as the non-relativistic
limit of the Kaluza-Klein framework. In \cite{duval} this approach was
applied to reformulate the equations of motions for particles in the
classical and quantum regimes in a covariant form. Hence, via the Eisenhart
lift we reveal the solution for the original dynamical system into the
solution for the equivalent system of geodesic equations. In the following
lines with describe lift in one-dimensional systems of the form of (\ref%
{ns.01}).

We define the momentum $p_{x}=\dot{x},$ thus, equation (\ref{ns.01}) is
expressed by the equivalent first-order dynamical system%
\begin{equation}
\dot{x}=p_{x}~,~\dot{p}_{x}=F\left( x\right) \text{,}  \label{ns.02}
\end{equation}%
The latter equations are the Hamilton's equation for the function%
\begin{equation}
H=\frac{1}{2}p_{x}^{2}+V\left( x\right) ~,~F\left( x\right) =-V\left(
x\right) _{,x}.  \label{ns.03}
\end{equation}

Hamiltonian (\ref{ns.03})\ describes the motion of a particle in the
one-dimensional space with line element $ds^{2}=dx^{2}$, under the potential 
$V\left( x\right) $. Function (\ref{ns.03}) is a conservation law for the
equations of motion (\ref{ns.02}) with value $H=h$.

\subsubsection{Equivalent Hamiltonian $H_{1+1}$}

We employ the Eisenhart lift and we introduce the new Hamiltonian function%
\begin{equation}
H_{1+1}\equiv \frac{1}{2}p_{x}^{2}+\frac{\alpha }{2}V\left( x\right)
p_{z}^{2}=h_{1+1}.  \label{ns.04}
\end{equation}%
This function describes the geodesic equations for the two-dimensional space
with line element%
\begin{equation}
ds_{\left( 1+1\right) }^{2}=dx^{2}+\frac{1}{\alpha V\left( x\right) }%
dz^{2}\,,  \label{ns.05}
\end{equation}%
which is known as the Eisenhart metric

The geodesic equations for the two-dimensional line element (\ref{ns.05}) are%
\begin{eqnarray}
\dot{x} &=&p_{x}~,~\dot{z}=\alpha V\left( x\right) p_{z}~,  \label{ns.06} \\
\dot{p}_{x} &=&-\frac{\alpha }{2}V_{,x}p_{z}^{2}~,~\dot{p}_{z}=0.
\label{ns.07}
\end{eqnarray}%
We observe that momentum $p_{z}=p_{z}^{0}$ is a conservation law for the
geodesic equations. Hence, by replacing in the rest of the equations we end
with the reduced system%
\begin{equation}
\dot{x}=p_{x}~,~\dot{p}_{x}=-\frac{\alpha }{2}V_{,x}\left( p_{z}^{0}\right)
^{2}~,~\dot{z}=\alpha V\left( x\right) p_{z}~.  \label{ns.07a}
\end{equation}

Hence, the original system (\ref{ns.02}) is recovered when $h_{1+1}=h$ and $%
\alpha \left( p_{z}^{0}\right) ^{2}=2$.

However, it is important to note that the extended Hamiltonian in the
Eisenhart lift is not uniquely defined. While Hamiltonian (\ref{ns.04})
represents the most common geometric description for the dynamical system (%
\ref{ns.01}), there exist alternative Hamiltonian functions as well.

At this point it is important to remark that the third equation in (\ref%
{ns.07a}), that is, for the extended variable $z$, plays no role in the
evolution of the original system. This dynamical system is foliated and the
set of initial conditions for the variable $z$,\ do not affect the evolution
of the original system in the variables $x,p_{x}$. That property remains
valid and for the equivalent Hamiltonians that we discuss bellow. 

\subsubsection{Equivalent Hamiltonian $H_{1+2}$}

Consider the extended Hamiltonian function

\begin{equation}
H_{1+2}\equiv \frac{1}{2}p_{x}^{2}+V\left( x\right)
p_{u}^{2}+p_{u}p_{v}=h_{n+2},  \label{ns.08}
\end{equation}%
with equations of motion%
\begin{eqnarray}
\dot{x} &=&p_{x}~,~\dot{u}=2V\left( x\right) p_{u}+p_{v}~,~\dot{v}=p_{u},
\label{ns.09} \\
\dot{p}_{x} &=&-V_{,x}p_{u}^{2}~,~\dot{p}_{u}=0~,~\dot{p}_{v}=0.
\label{ns.10}
\end{eqnarray}

Momentum $p_{u}$ and $p_{v}$ are conservation laws. By replacing in the rest
of the equations we end with the Hamiltonian system (\ref{ns.03}), assuming
that 
\begin{equation}
p_{u}^{2}=1~\text{and }p_{u}p_{v}-h_{1+2}=h.\ 
\end{equation}

The corresponding Eisenhart metric is described by the pp-wave spacetime 
\begin{equation}
ds_{\left( 1+2\right) }^{2}=dx^{2}+2dudv-2V\left( x\right) du^{2}.
\label{ns.11}
\end{equation}%
If we select $h_{1+2}=0$, then equations (\ref{ns.09}), (\ref{ns.10})
correspond to the null geodesics of space (\ref{ns.11}). As we shall discuss
in the following Section null geodesics have the property to be invariant
under conformal transformations. This characteristic has been applied before
to discover the original of the high-dimensional symmetry group in dynamical
systems \cite{duval2}. For possible extensions we refer the reader to the
recent discussion in \cite{zh1}.

The two lifts discussed thus far are commonly applied in the literature. The
first one is known as the Riemannian lift \cite{en4}, while the second one
is referred to as the Lorentzian lift \cite{en4}.

\subsection{New extended Hamiltonians}

We proceed with the introduction of two new lifts which, as we shall see,
can be applied to linearize equation (\ref{ns.01}) for specific functions of 
$F(x)$.

\subsubsection{Equivalent Hamiltonian $H_{1+3}$}

Assuming the extended Hamiltonian function

\begin{equation}
H_{1+3}\equiv \frac{1}{2}p_{x}^{2}+\frac{\alpha }{2}F_{1}\left( x\right)
p_{z}^{2}+F_{2}\left( x\right) p_{u}^{2}+p_{u}p_{v}=h_{n+3},  \label{h.12}
\end{equation}%
which is a combination of the Riemannian and Lorentzian lifts. The equations
of motion are given by 
\begin{eqnarray}
\dot{x} &=&p_{x}~,~\dot{z}=\alpha F_{1}\left( x\right) p_{z}~,~\dot{u}%
=2F_{2}\left( x\right) p_{u}+p_{v}~,~\dot{v}=p_{u}~, \\
\dot{p}_{x} &=&-\left( \frac{\alpha }{2}F_{1}\left( x\right)
_{,x}p_{z}^{2}+F_{2}\left( x\right) _{,x}p_{u}^{2}\right) ~, \\
\dot{p}_{z} &=&0~,~\dot{p}_{u}=0~,~\dot{p}_{v}=0.
\end{eqnarray}%
The conservation laws of the dynamical system with Hamiltonian (\ref{h.12})
are the momentum $p_{z},~p_{u}$ and $p_{v}$. The original Hamiltonian (\ref%
{ns.03}) is recovered when $V$%
\begin{equation}
V\left( x\right) =\frac{1}{2}\alpha F_{1}\left( x\right)
p_{z}^{2}+F_{2}\left( x\right) p_{u}^{2}+p_{u}p_{v}  \label{h.140}
\end{equation}%
and 
\begin{equation}
p_{u}p_{v}-h_{1+3}=h.
\end{equation}
Finally, the corresponding Eisenhart metric has the line element%
\begin{equation}
ds_{\left( 1+3\right) }^{2}=dx^{2}+\frac{1}{\alpha F_{1}\left( x\right) }%
dz^{2}+2dudv-2F_{2}\left( x\right) du^{2}.  \label{h.14}
\end{equation}

\subsubsection{Equivalent Hamiltonian $\hat{H}_{1+3}$}

We introduce the extended Hamiltonian function as follows 
\begin{equation}
\hat{H}_{1+3}=\frac{1}{2}p_{x}^{2}+\frac{1}{2}\alpha F_{1}\left( x\right)
p_{z}^{2}+V_{1}\left( x\right) p_{u}^{2}+V_{2}\left( x\right) p_{u}p_{v}=%
\hat{h}_{1+3}  \label{h.15}
\end{equation}%
with equations of motion%
\begin{eqnarray}
\dot{x} &=&p_{x}~,~\dot{z}=\alpha F_{1}\left( x\right) p_{z}~,~\dot{u}%
=2V_{1}\left( x\right) p_{u}+V_{2}\left( x\right) p_{v}~,~\dot{v}%
=V_{2}\left( x\right) p_{u}~, \\
\dot{p}_{x} &=&-\left( \frac{1}{2}\alpha F_{1}\left( x\right)
_{,x}p_{z}^{2}+V_{1}\left( x\right) _{,x}p_{u}^{2}+V_{2}\left( x\right)
_{,x}p_{u}p_{v}\right) ~, \\
\dot{p}_{z} &=&0~,~\dot{p}_{u}=0~,~\dot{p}_{v}=0.
\end{eqnarray}%
The corresponding Eisenhart metric is defined as 
\begin{equation}
d\hat{s}_{\left( 1+3\right) }^{2}=dx^{2}+\frac{1}{\alpha F_{1}\left(
x\right) }dz^{2}+\frac{2}{V_{2}\left( x\right) }dudv-2\frac{V_{1}\left(
x\right) }{\left( V_{2}\left( x\right) \right) ^{2}}du^{2}.  \label{h.16}
\end{equation}%
and the original Hamiltonian system (\ref{ns.03}) is recovered~for 
\begin{equation}
V\left( x\right) =\frac{1}{2}\alpha F_{1}\left( x\right)
p_{z}^{2}+V_{1}\left( x\right) p_{u}^{2}+V_{2}\left( x\right) p_{u}p_{v}
\end{equation}%
and $\hat{h}_{1+3}=h$.

Without loss of generality in the latter extension we can assume that $%
F_{1}\left( x\right) =1$, such that the original system to be recovered when 
\begin{equation}
V\left( x\right) =V_{1}\left( x\right) p_{u}^{2}+V_{2}\left( x\right)
p_{u}p_{v}
\end{equation}
and 
\begin{equation}
\frac{1}{2}\alpha p_{z}^{2}=h~\text{\ and }\hat{h}_{1+3}=0.
\end{equation}%
Hence, the latter dynamical systems to describe the null geodesics of the
line element (\ref{h.16}).

\section{Geometric Linearization of Newtonian systems}

\label{sec31}

In the previous section, we established the geometric representation of
equation (\ref{ns.01}). Therefore, the global linearization problem of
equation (\ref{ns.01}) is equivalent to the linearization of the
corresponding geodesic equations for each of the extended Hamiltonians.
Before search into the analysis of this problem, we first discuss the
application of conformal transformations for null geodesic equations.
Specifically, we focus on autonomous Hamiltonian systems where the value of
the Hamiltonian conservation law is constrained to be zero.

\begin{definition}
The two metric tensors $g_{ij}\left( x^{k}\right) $,~$\bar{g}_{ij}\left(
x^{k}\right) $ are conformally related if and only if there exist a set of
coordinates where $\bar{g}_{ij}=N^{2}\left( x^{k}\right) g_{ij}$. If $g_{ij}$
is the flat space, then $\bar{g}_{ij}$ is a conformally flat space.
\end{definition}

\begin{lemma}
Consider the Riemannian space with metric $g_{ij}\left( x^{k}\right) $ and
of dimension $n=\dim g_{ij}$. For $n=2$, the space $g_{ij}$ is always
conformally flat, and it is flat if the Ricci scalar is zero. When $n=3$,
the Riemannian space is conformally flat if and only if \ the Cotton-York
tensor is zero. The Cotton-York tensor is expressed in terms of the Riemann
tensor and of the Ricci scalar as follows 
\begin{equation}
C_{\mu \nu \kappa }=R_{\mu \nu ;\kappa }-R_{\kappa \nu ;\mu }+\frac{1}{4}%
\left( R_{;\nu }g_{\mu \kappa }-R_{;\kappa }g_{\mu \nu }\right) ,
\end{equation}%
Finally, for $n>3$, the space is conformally flat if and only if the Weyl
tensor is zero. The latter tensor is defined as 
\begin{equation}
C_{\mu \nu \kappa \lambda }=R_{\mu \nu \kappa \lambda }-\frac{2}{n-2}\left(
g_{\mu \lbrack \kappa }R_{\lambda ]\nu }-g_{\nu \lbrack \kappa }R_{\lambda
]\mu }\right) +\frac{2}{\left( n-1\right) \left( n-2\right) }Rg_{\mu \lbrack
\kappa }g_{\lambda ]\nu }.
\end{equation}
\end{lemma}

Consider now the geodesic Lagrangian for the Riemannian space with metric $%
g_{ij}$%
\begin{equation}
L\left( x^{k},\dot{x}^{k}\right) =\frac{1}{2}g_{ij}\left( x^{k}\right) \dot{x%
}^{i}\dot{x}^{j},
\end{equation}%
where $\dot{x}=\frac{dx}{dt}$. The Action reads%
\begin{equation}
S=\int dt\left( L\left( x^{k},\dot{x}^{k}\right) \right) =\int dt\left( 
\frac{1}{2}g_{ij}\dot{x}^{i}\dot{x}^{j}\right) .
\end{equation}

Under the change of variable $d\tau =N^{2}\left( x^{i}\right) dt,$we
introduce the conformal Lagrangian%
\begin{equation}
\bar{L}\left( x^{k},x^{\prime k}\right) =\frac{1}{2}N^{2}\left( x^{k}\right)
g_{ij}x^{\prime i}x^{\prime j},
\end{equation}%
in which $x^{\prime i}=\frac{dx}{d\tau }$.

The two Lagrangians $L\left( x^{k},\dot{x}^{k}\right) $ and $\bar{L}\left(
x^{k},x^{\prime k}\right) $ describe the geodesic equations for the two
conformal related metrics $g_{ij}\left( x^{k}\right) $ and~$\bar{g}%
_{ij}\left( x^{k}\right) $. For the geodesic equations of conformal
Lagrangians we have the following Theorem.

\begin{theorem}
The Euler-Lagrange equations for two conformal Lagrangians transform
covariantly under the conformal transformation relating the Lagrangians if
and only if the Hamiltonian vanishes.
\end{theorem}

This implies that null geodesics remain invariant under conformal
transformations. Therefore, the following corollary easily follows.

\begin{corollary}
The null geodesic equations of conformally flat Riemannian spaces can be
linearized.\ There exist a coordinate system where the geodesic equations
can be written in the form of the free particle.
\end{corollary}

In Appendix \ref{null} we demonstrate the application of the latter
corollary.

We consider the Eisenhart metrics for the corresponding extended Hamiltonian
functions $H_{1+1}$,~$H_{1+2}$,~$H_{1+3}$\ and $\hat{H}_{1+3}$.

For the $H_{1+1}$\ we impose that the energy density is not zero;
consequently, in order the geodesic equations to be linearized the metric (%
\ref{ns.05}) should be the flat space. Thus, the two-dimensional space (\ref%
{ns.05}) is the flat space if and only if the Ricci scalar is zero, i.e. 
\begin{equation}
2V_{,xx}V-3\left( V_{,x}\right) ^{2}=0,
\end{equation}%
with solution $V_{A}\left( x\right) =\frac{V_{0}}{\left( x-x_{0}\right) ^{2}}
$. Without loss of generality we assume $x_{0}=0$. 

Moreover, for the rest Hamiltonian functions $H_{1+2}$,~$H_{1+3}$\ and $\hat{%
H}_{1+3}$, we assume that they describe null geodesics. Hence, we require
the Eisenhart metrics (\ref{ns.11}), (\ref{h.14}) and (\ref{h.16}) to be
conformally flat. 

Thus, the three-dimensional line element (\ref{ns.11}) describes a
conformally flat geometry if and only if 
\begin{equation}
V_{,xxx}=0
\end{equation}%
from where it follows $V_{B}\left( x\right) =\frac{\omega }{2}\left(
x-x_{0}\right) ^{2}+\omega _{0}$. 

Furthermore, the four-dimensional line element (\ref{h.14}) describes a
conformally flat geometry when the free functions $F_{1}\left( x\right) $,~$%
F_{2}\left( x\right) $\ are constraint by the following system of
differential equations,%
\begin{eqnarray}
2F_{1,xx}F-3\left( F_{1,x}\right) ^{2} &=&0, \\
2F_{2,xx}F_{1}+F_{2,x}F_{1,x} &=&0.
\end{eqnarray}%
The solution of the latter system is $F_{1}\left( x\right) =\frac{1}{\left(
F_{1}^{0}x-F_{1}^{1}\right) ^{2}}$,~$F_{2}\left( x\right)
=F_{2}^{0}+F_{2}^{1}\left( x-\frac{F_{1}^{1}}{F_{1}^{0}}\right) ^{2}$.
Hence, by replacing in\ (\ref{h.140}) and eliminate the non-essential
constants we end with the Ermakov potential with oscillator $V_{C}\left(
x\right) =\frac{\omega }{2}x^{2}+\frac{V_{0}}{x^{2}}$.

Finally, for the line element (\ref{h.16}) we assume without loss of
generality that $F_{1}\left( x\right) =1$, hence, the space is conformally
flat when 
\begin{eqnarray}
3V_{2}\left( V_{1,xx}V_{2}-3V_{1,x}V_{2,x}\right) +6V_{1}\left(
V_{2,x}\right) ^{2} &=&0, \\
V_{2,xx}V_{2}-\left( V_{2,x}\right) ^{2} &=&0.
\end{eqnarray}%
Hence, $V_{1}\left( x\right) =V_{1}^{0}e^{\lambda x}+V_{1}^{1}e^{2\lambda x}$%
\ and $V_{2}\left( x\right) =V_{2}^{0}e^{\lambda x}$. Consequently, the
Morse potential $V\left( x\right) =V_{1}^{0}e^{\lambda
x}+V_{2}^{0}e^{2\lambda x}$\ follows.

The results are summarized in the following theorem.

\begin{theorem}
\label{theom} The Newtonian system (\ref{ns.01}), with $F\left( x\right)
=-V\left( x\right) _{,x}$ can be written as the free particle $\frac{d^{2}X}{%
dT^{2}}=0$, for the potentials: (A) Ermakov potential $V_{A}\left( x\right) =%
\frac{V_{0}}{x^{2}}$;~(B) the oscillator $V_{B}\left( x\right) =\frac{\omega 
}{2}x^{2}$; (C) the Ermakov potential with oscillator $V_{C}\left( x\right) =%
\frac{\omega }{2}x^{2}+\frac{V_{0}}{x^{2}}$ and (D) the Morse potential $%
V\left( x\right) =V_{1}^{0}e^{\lambda x}+V_{2}^{0}e^{2\lambda x}$, via the
Eisenhart lift and the equivalent Hamiltonians $H_{1+1}$,~$H_{1+2}$,~$H_{1+3}
$ and $\hat{H}_{1+3}$ $\ $respectively.
\end{theorem}

The proof of the latter theorem is presented in Appendix \ref{proof}. Let us
now demonstrate the latter result with some applications.

Consider the Hamiltonian $H_{1+1}$ for the Ermakov potential~$V_{A}\left(
x\right) =\frac{V_{0}}{x^{2}}$. That is, 
\begin{equation}
H_{1+1}\equiv \frac{1}{2}p_{x}^{2}+\frac{\alpha V_{0}}{2x^{2}}p_{z}^{2},
\label{h.20}
\end{equation}%
We introduce the new variables \cite{pal1s}%
\begin{equation}
x=\sqrt{X^{2}+Y^{2}}~,~z=\frac{1}{\sqrt{\alpha V_{0}}}\arctan \left( \frac{Y%
}{X}\right) ,
\end{equation}%
then, Hamiltonian (\ref{h.20}) becomes%
\begin{equation}
H_{1+1}=\frac{1}{2}\left( p_{X}^{2}+p_{Y}^{2}\right) ,  \label{h.21}
\end{equation}%
which is the Hamiltonian for the free particle. In a similar way we can
linearize and the rest Eisenhart metrics.

The solution of \ Hamilton's equations (\ref{h.21}) reveal four integration
constants, while the original system process only two integration constants.
Due to the constraints $h_{1+1}=h$\ and $\alpha \left( p_{z}^{0}\right)
^{2}=2$, we end with one free integration constant. As we discuss before
that third integration constant is related with parameter $z$\ and plays no
role in the solution for the original dynamical system. 

Assume now the Hamiltonian function $H_{1+2}$ with the quadratic potential $%
V_{B}\left( x\right) =\frac{\omega }{2}x^{2}$, i.e.%
\begin{equation}
H_{1+2}\equiv \frac{1}{2}p_{x}^{2}+\frac{\omega }{2}%
x^{2}p_{u}^{2}+p_{u}p_{v}.
\end{equation}%
Under the action of the point transformation%
\begin{equation}
x=\frac{X}{\sqrt{1+\bar{Z}^{2}}}~,~u=\frac{1}{\sqrt{\omega }}\arctan \bar{Z}%
~,~v=\sqrt{\omega }\left( 2Y-\bar{Z}+\frac{\bar{Z}}{\left( 1+\bar{Z}%
^{2}\right) }R^{2}\right) ~~,~\bar{Z}=Y+Z
\end{equation}%
we end up with the Hamiltonian%
\begin{equation}
H_{1+2}=\left( 1+\bar{Z}^{2}\right) \left(
p_{X}^{2}+p_{Y}^{2}-p_{Z}^{2}\right) .
\end{equation}%
Transformation $\left( x,u\right) \rightarrow \left\{ X,\bar{Z}\right\} $ is
nothing else than the usual transformation provided by the Lie symmetry
analysis which relates the oscillator to the free particle \cite{Leach80a}.

We continue our analysis by introducing the Hamiltonian function 
\begin{equation}
\hat{H}_{1+3}=\frac{1}{2}p_{x}^{2}+\frac{1}{2}p_{z}^{2}+V_{10}e^{\lambda
x}p_{u}^{2}+V_{20}e^{\lambda x}p_{u}p_{v}
\end{equation}%
which leads to the Newtonian system (\ref{ns.01}) for the exponential
potential $V\left( x\right) =V_{1}^{0}e^{\lambda x}$. Under the change of
variables 
\begin{equation}
x=\frac{2}{\lambda }\ln \left( \frac{\lambda \sqrt{V_{10}Z^{2}+V_{20}\left(
WX-YZ\right) }}{\sqrt{2}V_{20}W}\right) ~,~z=-\frac{2i}{\lambda }\ln \left( 
\frac{\lambda \sqrt{V_{10}Z^{2}+V_{20}\left( WX-YZ\right) }}{\sqrt{2}V_{20}W}%
\right) ,
\end{equation}%
\begin{equation}
u=\frac{Z}{W}~,~v=\frac{Y}{W},
\end{equation}%
the latter Hamiltonian function reads%
\begin{equation}
\hat{H}_{1+3}=e^{-2i\lambda z}\left(
-V_{20}p_{X}p_{W}+2V_{10}p_{Y}^{2}+2V_{20}p_{Y}P_{Z}\right) =0,
\end{equation}%
where the resulting equations of motion are linear, that is,%
\begin{equation}
\ddot{X}=0~,~\ddot{Y}=0~,~\ddot{Z}=0~,\ddot{W}=0\text{.}
\end{equation}

\section{Conclusions}

\label{sec4}

We conducted a study on the linearization of the equation of motion for a
particle moving along a line under the influence of an autonomous force,
within the framework of a Hamiltonian dynamical system. To achieve this, we
employed the strategy of the Eisenhart lift to formulate a system of
geodesic equations that are equivalent to the original problem. By defining
various sets of equivalent geodesic Hamiltonian systems and working within
the framework of Riemannian geometry, we were able to determine the
functional forms for the force term. This allowed us to identify cases where
the geodesic equations could be linearized, effectively transforming them
into the equations of motion for a free particle in a plane geometry.

Our investigation revealed that for the Ermakov potential, the oscillator,
and the Morse potential, we could define a higher-dimensional geometry where
the equivalent geodesic equations could be linearized. Consequently, for
these potentials, the Newtonian system under our consideration becomes
equivalent to that of a free particle, under a change of variables.

This method opens up new directions for studying classical systems and for
the linearization of differential equations. In future work, we aim to
extend this analysis to higher-dimensional dynamical systems.

\bigskip

\textbf{Data Availability Statements:} Data sharing is not applicable to
this article as no datasets were generated or analyzed during the current
study.

\bigskip

\textbf{Acknowledgements: }The author thanks the support of Vicerrector\'{\i}%
a de Investigaci\'{o}n y Desarrollo Tecnol\'{o}gico (Vridt) at Universidad
Cat\'{o}lica del Norte through N\'{u}cleo de Investigaci\'{o}n Geometr\'{\i}%
a Diferencial y Aplicaciones, Resoluci\'{o}n Vridt No - 096/2022 and Resoluci%
\'{o}n Vridt No - 098/2022. Part of this study was supported by FONDECYT
1240514.

\appendix

\section{One-dimensional constraint system}

\label{null}

Consider the null geodesics for the conformally flat line element (\ref%
{ns.05}). We shall see that for arbitrary potential the dynamical system can
be global linearized.

We introduce the change of variables $dx=F\left( X\right) dX$, thus, the
line element is 
\begin{equation}
ds_{\left( 1+1\right) }^{2}=F^{2}\left( X\right) dX^{2}+\frac{1}{\alpha
V\left( \int F\left( X\right) dX\right) }dz^{2},
\end{equation}%
we select function $F\left( X\right) $ such that$\,F^{2}\left( X\right) =%
\frac{1}{V\left( \int F\left( X\right) dX\right) }$. Consequently,
Hamiltonian (\ref{ns.04}) for the null geodesics reads%
\begin{equation}
H_{1+1}\equiv V\left( \int F\left( X\right) dX\right) \left( p_{X}^{2}+\frac{%
1}{\alpha }p_{z}^{2}\right) =0.  \label{ns.20}
\end{equation}%
with equations of motion 
\begin{eqnarray}
\dot{X} &=&V\left( \int F\left( X\right) dX\right) p_{X}~,~\dot{z}=V\left(
\int F\left( X\right) dX\right) p_{z}~, \\
\dot{p}_{X} &=&V\left( \int F\left( X\right) dX\right) _{,X}\left( p_{X}^{2}+%
\frac{1}{\alpha }p_{z}^{2}\right) ~,~\dot{p}_{Z}=0.
\end{eqnarray}%
With the use of the constraint (\ref{ns.20}) it follows%
\begin{eqnarray}
\dot{X} &=&V\left( \int F\left( X\right) dX\right) p_{X}~,~\dot{z}=V\left(
\int F\left( X\right) dX\right) p_{z}~, \\
\dot{p}_{X} &=&0~,~\dot{p}_{Z}=0.
\end{eqnarray}

We consider the new independent variable $d\tau =V\left( \int F\left(
X\right) dX\right) dt$, from where it follows%
\begin{equation}
X^{\prime }=p_{X}~,~z^{\prime }=p_{z}~,~p_{X}^{\prime }=0~,~p_{z}^{\prime
}=0,
\end{equation}%
or in the equivalently form%
\begin{equation}
X^{\prime \prime }=0~,~z^{\prime \prime }=0.
\end{equation}

\end{document}